\documentclass[11pt,a4paper]{article}
 
\usepackage{amsmath,amsthm,amssymb}

\usepackage{cite}

\pdfoutput=1

\usepackage{graphics,graphicx}
\usepackage{epsfig}
\usepackage{multicol}
\usepackage{color}
\makeatletter
\@addtoreset{equation}{section}
\makeatother

\usepackage{braket}
\usepackage{bm}

\begin{document} 
\sloppy

\begin{center}
\LARGE{{\bf Dimensionally-dependent uncertainty relations, or why we (probably) won't see micro-black holes at the LHC, even if large extra dimensions exist}}
\end{center}

\begin{center}
\large{Matthew J. Lake${}^{a,b,c,d,e*}$\footnote{matthewjlake@narit.or.th}, Shi-Dong Liang${}^{c,f}$\footnote{stslsd@mail.sysu.edu.cn} and Anucha Watcharapasorn${}^{b,g}$\footnote{anucha@stanfordalumni.org}} 
\end{center}
\begin{center}
\emph{$^{a}$National Astronomical Research Institute of Thailand, \\ 260 Moo 4, T. Donkaew,  A. Maerim, Chiang Mai 50180, Thailand \\}
\emph{$^{b}$Department of Physics and Materials Science, \\ Faculty of Science, Chiang Mai University, \\ 239 Huaykaew Road, T. Suthep, A. Muang, Chiang Mai 50200, Thailand \\}
\emph{$^{c}$School of Physics, Sun Yat-Sen University, \\ Guangzhou 510275, People’s Republic of China \\}
\emph{$^{d}$Department of Physics, Babe\c s-Bolyai University, \\ Mihail Kog\u alniceanu Street 1, 400084 Cluj-Napoca, Romania \\}
\emph{$^{e}$Office of Research Administration, Chiang Mai University, \\ 239 Huaykaew Rd, T. Suthep, A. Muang, Chiang Mai 50200, Thailand \\}
\emph{$^{f}$State Key Laboratory of Optoelectronic Material and Technology, \\ Guangdong Province Key Laboratory of Display Material and Technology, 
\\ Sun Yat-Sen University, Guangzhou 510275, People's Republic of China \\}
\emph{$^{g}$Center of Excellence in Quantum Technology, \\ Faculty of Engineering, Chiang Mai University, \\ 239 Huaykaew Rd, T. Suthep, A. Muang, Chiang Mai 50200, Thailand}
\vspace{0.1cm}
\end{center}

\begin{abstract}
We present a simple gedanken experiment in which a compact object traverses a spacetime with three macroscopic spatial dimensions and $n$ compact dimensions. 
The compactification radius is allowed to vary, as a function of the object's position in the four-dimensional space, and we show that the conservation of gravitational self-energy implies the dimensional dependence of the mass-radius relation. 
In spacetimes with extra dimensions that are compactified at the Planck scale, no deviation from the four-dimensional result is found, but, in spacetimes with extra dimensions that are much larger than the Planck length, energy conservation implies a deviation from the normal Compton wavelength formula. 
The new relation restores the symmetry between the Compton wavelength and Schwarzschild radius lines on the mass-radius diagram and precludes the formation of black holes at TeV scales, even if large extra dimensions exist. 
We show how this follows, intuitively, as a direct consequence of the increased gravitational field strength at distances below the compactification scale. 
Combining these results with the heuristic identification between the Compton wavelength and the minimum value of the position uncertainty, due to the Heisenberg uncertainty principle, suggests the existence of generalised, higher-dimensional uncertainty relations. 
These relations may be expected to hold for self-gravitating quantum wave packets, in higher-dimensional spacetimes, with interesting implications for particle physics and cosmology in  extra-dimensional scenarios.
\end{abstract}

\section*{}
{{\bf Keywords}: compactification, higher dimensions, Compton wavelength, primordial black holes, generalised uncertainty relations, self-gravity}




\section{Introduction} \label{Sec.1}

For over forty years, models with compact extra dimensions have attracted a great deal of attention in the theoretical physics literature. 
Much of this interest was motivated by superstring theory, which is only consistent in ten spacetime dimensions \cite{Green:1987sp,Green:1987mn}, requiring six space-like dimensions to be curled up on scales that make them inaccessible to current high-energy experiments. 
Theoretically, the compactification scale may be as low as the Planck length, placing it forever beyond the reach of terrestrial particle physics, but models with effective compactification scales as high as a millimetre have also been proposed \cite{Arkani-Hamed:1998jmv,Antoniadis:1998ig}. 
Prior to the start-up of the Large Hadron Collider (LHC), in 2010, interest in the phenomenology of higher-dimensional models reached an all-time high. 
It peaked again following beam upgrades in 2015, but, since then, has been in decline.

In the heady days of the late nineteen-nineties and the first two decades of the twenty-first century, it was hoped, and, indeed, argued persuasively in the scientific literature, that the TeV scale experiments soon to be conducted at CERN would enable the direct detection of compact dimensions with length scales down to $\sim 10^{-19}$ m. 
It was claimed that these, so-called `large' extra dimensions, could induce the formation of microscopic black holes \cite{Arkani-Hamed:1998sfv,Bleicher:2011uj,Kiritsis:2011qv,CMS:2010oej,Park:2012fe,Bellagamba:2012wz,Mureika:2011hg,Taliotis:2012sx,Winstanley:2013dua,Nicolini:2013ega,Torres:2013kda,Alberghi:2013hca,Belyaev:2014ljc,Hou:2015gba,Sokolov:2016lba}, also known as primordial black holes (PBH), in reference to their cosmic cousins \cite{Carr:2005zd,Green:2020jor,Carr:2020xqk,Carr:2020gox,Escriva:2022duf,Friedlander:2022ttk}. 
These claims even attracted considerable attention in the popular press \cite{APS,NYT,NASA,BBC,Huffpost,Forbes}.

The argument behind this assertion was straightforward and reasonable. 
It is well known that the radius of an uncharged and non-spinning (Schwarzschild) black hole depends, not only on its mass, but also on the dimensionality of the spacetime it inhabits. 
The higher-dimensional Schwarzschild radius varies as $\mathcal{R}_{\rm S} \propto M^{\frac{1}{1+n}}$, where $n$ is the number of space-like extra dimensions, over and above the three Hubble scale dimensions that make up the macroscopic Universe \cite{Weinberg,Horowitz:2012nnc}. 
Thus, assuming that the usual mass-dependence of the Compton wavelength, $R_{\rm C} \propto M^{-1}$, remains unchanged in the presence of the compact space, the intersection between 
$\mathcal{R}_{\rm S}$ and $R_{\rm C}$ occurs close to the critical values
\begin{eqnarray} \label{HD_Planck} 
R_{\rm crit} = \left(\frac{\hbar G_{4+n}}{c^3}\right)^{\frac{1}{2+n}}  \, , \ M_{\rm crit} = \left(\frac{\hbar^{1+n}c^{1-n}}{G_{4+n}}\right)^{\frac{1}{2+n}} \, .
\end{eqnarray} 
For $n \geq 0$, these expressions serve as the definitions of the Planck length, and mass, respectively \cite{Horowitz:2012nnc,Maartens:2010ar}. 
Since, in spacetimes with $n$ compact dimensions, the four-dimensional Newton's constant is related to its higher-dimensional counterpart, and to the compactification radius $R_{\rm E}$, via \cite{Maartens:2010ar} 
\begin{eqnarray} \label{G_4:G_4+n}
G_{4+n} = G_{4}R_{\rm E}^{n} \, ,
\end{eqnarray}
it follows that, for sufficiently large $R_{\rm E}$, the mass-energy needed to create a black hole may be brought within the TeV range of the LHC. 

More recently, new phenomenological models have been proposed, in which the possible dimensional dependence of the Compton wavelength has been explored \cite{Carr:2017aws,Lake:2018hyv,Carr:2022ndy,Lake:2016enn}, via the so-called black hole--uncertainty principle (BHUP) correspondence, which is also referred to as the Compton--Schwarzschild correspondence in the literature \cite{Carr:2011pr,Carr:2015nqa,Carr:2014mya,Lake:2015pma,Singh:2017wrb,Singh:2017ipg,daSilva:2022xgx}. 
This modification alters the intersection with the Schwarzschild radius, and is capable of restoring complete symmetry to the $(M,R)$ diagram, pushing the threshold for black hole formation back up to the four-dimensional Planck mass, $M_{\rm Pl} = \sqrt{\hbar c/G_4}$. 
However, despite the various arguments used to justify these models \cite{Carrr:2017aws,Lake:2018hyv,Carr:2022ndy,Lake:2016enn}, the proposed dimensional dependence lacks a clear physical motivation. 
In this work, we motivate them in a more direct way, by outlining a clear physical mechanism that is capable of altering the mass-radius relation of any compact object, including that of fundamental particles.  

The structure of this paper is as follows. 
In the main body of the work, Sec. \ref{Sec.2}, we present a simple gedanken experiment in a hypothetical universe with three macroscopic spatial dimensions and $n$ compact extra dimensions. 
The compactification radius is allowed to vary as a function of position in the four-dimensional subspace, which is divided into three regions. 
In the first region, the extra dimensions are compactified at the four-dimensional Planck-scale, $R_{\rm Pl} = \sqrt{\hbar G/c^3}$, while in the third they are compactified at a much larger radius, $R_{\rm E} > R_{\rm Pl}$. 
The second region, in which the compactification scale grows monotonically, interpolates smoothly between the other two. 
We then consider a compact object, which passes from region 1 to region 3, and impose the conservation of gravitational self-energy. 
Roughly speaking, since gravity becomes stronger on scales $R_{\rm Pl} < R < R_{\rm E}$, as we move through region 2, the radius of the object must increase, in order to keep its gravitational self-energy constant. 
Furthermore, since rest mass is conserved during this transition, it follows that the mass-radius relation must be modified. 

In this study, we perform explicit calculations by assuming that the gravitational potential of the object can be approximated by the weak field (Newtonian) limit. 
However, despite this, our analysis correctly reproduces well known results for strongly-gravitating objects, such as higher-dimensional black holes and neutron stars, up to numerical factors of order unity, which is consistent with the non-relativistic approximation. 
This gives us confidence in the method, which we then extend to the study of fundamental particles, for which the non-relativistic approximation is undoubtedly valid.

We verify that, beginning with an effectively four-dimensional black hole in region 1, we obtain the correct (order of magnitude) expression for the higher-dimensional Schwarzschild radius in region 3. 
This gives us confidence in our procedure, which we note is agnostic to the initial mass-radius relation of the object. 
We then consider a fundamental particle, by beginning instead with the standard formula for the Compton wavelength, and obtain an effective, higher-dimensional Compton radius, in the third region. 
Its implications for the (non-)formation of black holes at the LHC, as well as for the quantum mechanical uncertainty relations of self-gravitating wave packets in higher-dimensional spacetimes, are briefly discussed. 
We summarise our conclusions, and consider the prospects for future work on this model, in Sec. \ref{Sec.3}. 

\section{The gedanken experiment} \label{Sec.2}

Let us assume, for simplicity, that the compact object we consider is spherically symmetric. 
In region 1, its internal energy is, therefore
\begin{eqnarray} \label{EnergyRegion1}
E = Mc^2 - \alpha_4 \frac{G_4M^2}{R} \, ,
\end{eqnarray}
where $\alpha_4$ is a numerical constant determined by the mass profile of the sphere, $M(r)$. 
For example, $\alpha_4 = 3/5$ for a sphere of uniform density and should be of order unity for all non-pathological profiles \cite{GravEnergySphere}.
Here, $R$ denotes the effective macrosopic radius of the object and Eq. (\ref{EnergyRegion1}) holds for all $R \gtrsim R_{\rm E} = R_{\rm Pl}$. 
We note that distances below this scale cannot be probed directly, by either black holes or fundamental particles, due to the intersection of the Compton wavelength and Schwarzschild radius lines near the Planck point on the $(M,R)$ diagram \cite{Carr:2014mya,Lake:2015pma}. 

Requiring $E \leq 0$, which implies a bound state, yields
\begin{eqnarray} \label{E<0_Region1}
R \leq \alpha_4 \frac{G_4M}{c^2} \, .  
\end{eqnarray}
For $\alpha_4 = 2$, we then recover the condition 
\begin{eqnarray} \label{SchwarzschildRadius_Region1}
R \leq R_{\rm S}(M) = \frac{2G_4M}{c^2} \, , 
\end{eqnarray}
where $R_{\rm S}$ is the four-dimensional Schwarzschild radius. 
Thus, if Eq. (\ref{SchwarzschildRadius_Region1}) is satisfied, the object is a black hole in the first region. 
For $\alpha_4 = 9/4$, an analogous condition implies violation of the Buchdahl inequality \cite{Buchdahl:1959zz} and the sphere may be viewed as a compact star undergoing collapse. 
Conversely, for $E > 0$, $R > (9/4)G_{4}M/c^2$, the object is stable against its own self-gravity. 

Setting $R = R_{\rm C}$, where 
\begin{eqnarray} \label{ComptonRadius_Region1}
R_{\rm C}(M) = \frac{\hbar}{Mc}
\end{eqnarray}
is the standard Compton radius \cite{Compton}, Eq. (\ref{EnergyRegion1}) implies that a fundamental particle is stable against gravitational collapse ($E > 0$) when
\begin{eqnarray} \label{Planck_limits}
M \lesssim M_{\rm Pl} \, , \quad R_{\rm C}(M) \gtrsim R_{\rm Pl} \, ,
\end{eqnarray}
where 
\begin{eqnarray} \label{4D_Planck}
R_{\rm Pl} = \sqrt{\frac{\hbar G_{4}}{c^3}} \simeq 10^{-35} \, {\rm m} \, , \quad M_{\rm Pl} = \sqrt{\frac{\hbar c}{G_{4}}} \simeq 10^{-8} \, {\rm kg} \, .
\end{eqnarray}
Equation (\ref{Planck_limits}) justifies our previous assertion that Eq. (\ref{EnergyRegion1}) holds, for $R \gtrsim R_{\rm Pl}$, when the extra dimensions are compactified at the (four-dimensional) Planck scale. 
For fundamental particles, this corresponds to the region $M \lesssim M_{\rm Pl}$, whereas, for black holes, it corresponds to 
\begin{eqnarray} \label{Schwarz_limits}
M \gtrsim M_{\rm Pl} \, , \quad R_{\rm S}(M) \gtrsim R_{\rm Pl} \, .
\end{eqnarray}
The intersection of the standard Compton line and the four-dimensional Schwarzschild line near the Planck point then precludes the existence of {\it any} fundamental object with $R(M) \lesssim R_{\rm Pl}$. 

In the third region, the internal energy of the object is given by Eq. (\ref{EnergyRegion1}), for $R > R_{\rm E}$, where $R_{\rm E} > R_{\rm Pl}$ is the compactification radius, but by 
\begin{eqnarray} \label{EnergyRegion3}
E = Mc^2 - \alpha_{4+n}\frac{G_{4+n}M^2}{\mathcal{R}^{1+n}} 
\end{eqnarray}
for $R_{\rm Pl} \leq \mathcal{R} \leq R_{\rm E}$. 
Here, $\mathcal{R}$ denotes the $(4+n)$-dimensional radius in region 3 and $\alpha_{4+n}$ is a numerical constant determined by the mass profile of the object in the higher-dimensional space. 
For simplicity, we assume that all $n$ extra dimensions are compactified on the same scale.   
The relation between $G_4$ and the higher-dimensional Newton's constant, $G_{4+n}$, is given by Eq. (\ref{G_4:G_4+n}) \cite{Maartens:2010ar}.

By choosing appropriate values of $\alpha_{4+n}$, we may recover the $(4+n)$-dimensional analogues of the Buchdahl bound \cite{Burikham:2015nma,Burikham:2015sro} and the Schwarzschild radius \cite{Horowitz:2012nnc}, from the energy conditions $E <(>) \, 0$. 
In any number of dimensions, the Buchdahl radius is proportional to the Schwarzschild radius, and, neglecting numerical factors of order unity, the latter may be written as 
\begin{eqnarray} \label{SchwarzschildRadius_Region3}
\mathcal{R}_{\rm S}(M) \simeq \left(\frac{G_{4+n}M}{c^2}\right)^{\frac{1}{1+n}} \simeq (R_{\rm S}(M)R_{\rm E}^n)^{\frac{1}{1+n}} \, , 
\end{eqnarray}
where $R_{\rm S}(M)$ again denotes the four-dimensional Schwarzschild radius, as in Eq. (\ref{SchwarzschildRadius_Region1}). 

Let us now consider a non-relativistic, self-gravitating sphere, with arbitrary mass-radius relation, passing from region 1 to region 3. 
Furthermore, let us assume that, whatever its mass-radius relation in the  four-dimensional space of the first region, the sphere remains small enough to be effectively $(4+n)$-dimensional in the third. 
Thus, in region 1, its radius in the three macroscopic spatial dimensions is $R(M) \gtrsim R_{\rm Pl}$ and, in region 3, its higher-dimensional radius satisfies $R_{\rm Pl} \lesssim \mathcal{R}(M) \lesssim R_{\rm E}$. 
If its internal energy remains unchanged, energy conservation then implies
\begin{eqnarray} \label{EnergyConservation}
\mathcal{R}(M) \simeq (R(M)R_{\rm E}^n)^{\frac{1}{1+n}} \, , 
\end{eqnarray}
again ignoring numerical factors of order unity, which is consistent with the non-relativistic approximation. 
Note that we again use the calligraphic font, $\mathcal{R}$, to denote radii in $(4+n)$ dimensions, and the normal font $R$ to denote four-dimensional radii. 

Substituting $R(M) \simeq R_{\rm S}(M)$ (\ref{SchwarzschildRadius_Region1}) into (\ref{EnergyConservation}), we recover the correct expression for the higher-dimensional Schwarzschild radius, $\mathcal{R}_{\rm S}(M)$ (\ref{SchwarzschildRadius_Region3}). 
Next, we note that, if $R_{\rm Pl} < \mathcal{R}_{\rm S}(M) < R_{\rm E}$, then $R_{\rm Pl} < R_{\rm S}(M) < R_{\rm E}$. 
It follows, immediately, that $\mathcal{R}_{\rm S}(M) > R_{\rm S}(M)$. 
This result can be understood intuitively as follows. 
Since, in the third region, the gravitational force is stronger than in the first on scales $\mathcal{R} < R_{\rm E}$, the radius of the black hole can neither decrease, nor remain constant, without increasing its internal energy. 
If this energy is conserved, the black hole must increase in size and the $(4+n)$-dimensional Schwarzschild radius, $\mathcal{R}_{\rm S}(M)$, must be larger than the four-dimensional radius, $R_{\rm S}(M)$. 
The relation between the two is fixed, by energy conservation, according to Eq. (\ref{SchwarzschildRadius_Region3}), 

Clearly, we may repeat a similar argument for stable compact objects obeying the four-dimensional Buchdahl bound in region 1. 
The same compact spheres then obey the higher-dimensional Buchdahl bound in region 3. 
Hence, although the argument presented above is simple and heuristic, it allows us to recover the same relations (to within an order of magnitude) as those obtained by exactly solving the gravitational field equations in $(4+n)$-dimensional spacetime \cite{Horowitz:2012nnc,Burikham:2015nma,Burikham:2015sro}. 

However, its greatest advantage is that it is agnostic to the mass-radius relation of the compact object. 
We may therefore apply it to fundamental particles, as well as to black holes and conventional fluid spheres. 
Thus, substituting $R(M) = R_{\rm C}(M) \propto M^{-1}$ (\ref{ComptonRadius_Region1}) into Eq. (\ref{EnergyConservation}), we obtain the higher-dimensional Compton wavelength,
\begin{eqnarray} \label{HD_Compton}
\mathcal{R}_{\rm C}(M) \simeq (R_{\rm C}(M)R_{\rm E}^n)^{\frac{1}{1+n}} 
\simeq R_{*}\left(\frac{M_{\rm Pl}}{M}\right)^{\frac{1}{1+n}}  \, , 
\end{eqnarray}
where 
\begin{eqnarray} \label{R_*}
R_{*} = (R_{\rm Pl}R_{\rm E}^{n})^{\frac{1}{1+n}} \, , 
\end{eqnarray}
so that $R_{\rm Pl} < R_{*} < R_{\rm E}$. 
It may be verified that the $(4+n)$-dimensional Compton and Schwarzschild lines intersect at the point $(M,R) \simeq (M_{\rm Pl},R_{*})$, so that the production of PBHs still requires energies of the order of the Planck energy \cite{Carrr:2017aws,Lake:2018hyv,Carr:2022ndy,Lake:2016enn}. 

This result can also be understood, intuitively, in the same way as our heuristic derivation of the higher-dimensional Schwarzschild radius. 
Namely, if the rest mass of the particle remains constant as it traverses the path from region 1 to region 3, its radius cannot remain constant, or decrease, without increasing its gravitational binding energy. 
Therefore, if its total internal energy remains constant, its radius must expand as it enters the higher-dimensional region, in which gravity is stronger, on scales $\mathcal{R} < R_{\rm E}$, than in four-dimensional space. 
Clearly, this relation must also hold for particles that were always confined to region 3.

To aid visualisation, a schematic representation of the gedanken experiment set up is given in Fig. 1. 
In Fig. 2a, the key length and mass scales of the standard scenario, corresponding to Eqs. (\ref{HD_Planck}), are depicted on the $(M,R)$ diagram, while the key scales for our scenario are depicted in Fig. 2b. 
The important difference between the two scenarios is that the former does not account for the self-gravitational energy of the particle, whereas the latter does, to within the accuracy permitted by the non-relativistic, weak-field approximation, which we also apply to micro-black holes. 
Maintaining this approximation, we may apply the usual, heuristic identification between the Compton wavelength formula and the limiting values of the Heisenberg uncertainty principle (HUP), 
\begin{eqnarray} \label{HUP_correspondence}
(\Delta X)_{\rm min} \simeq \mathcal{R}_{\rm C}(M) \, , \quad (\Delta P)_{\rm max} \simeq Mc \, , 
\end{eqnarray}
giving
\begin{eqnarray} \label{HD_HUP}
\Delta X \gtrsim R_{*}\left(\frac{M_{\rm Pl}c}{\Delta P}\right)^{\frac{1}{1+n}}  \, .
\end{eqnarray} 
We recall that, for $\Delta P \gtrsim Mc$, fundamental particles have sufficient energy to undergo pair-production, in interactions that conserve the relevant quantum numbers \cite{Peskin:1995ev,StandardModel}, yielding the limits in Eq. (\ref{HUP_correspondence}). 
These, in turn, correspond to the dimensionally-dependent uncertainty relation, Eq. (\ref{HD_HUP}).

Equation (\ref{HD_HUP}) may be expected to hold for self-gravitating wave packets, on scales $\mathcal{R} < R_{\rm E}$, in spacetimes with compact extra dimensions. 
By contrast, on scales  $\mathcal{R} > R_{\rm E}$, or when $R_{\rm E} \simeq R_{\rm Pl}$, the standard HUP,
\begin{eqnarray} \label{HUP}
\Delta X \gtrsim \frac{R_{\rm Pl}M_{\rm Pl}c}{\Delta P}  \, ,
\end{eqnarray} 
still holds, where we have rewritten $\hbar = R_{\rm Pl}M_{\rm Pl}c$. 

Finally, before concluding this section, we note that, although Eq. (\ref{HD_HUP}) represents a form of generalised uncertainty principle, which is valid for self-gravitating objects in higher-dimensional spacetimes, this is not the same as the  `generalised uncertainty principle' (GUP), commonly referred to in the quantum gravity literature (see, for example, \cite{Adler:2001vs,Maziashvili:2005pp,Sakalli:2022xrb,Xiang:2009yq,Lake:2018zeg,Lake:2023lvh} and references therein). 
In fact, the derivation of Eq. (\ref{HD_HUP}) is based on two fundamental assumptions, namely, (a) that the gravitational self-energy of the quantum wave packet is conserved in the presence of extra dimensions, and (b) that the standard HUP holds in their absence.

By contrast, the usual GUP is derived, via a gedanken experiment in four-dimensional spacetime, by considering the gravitational interaction between a measured particle and a probing photon. 
This gives rise to a correction term, to the position uncertainty $\Delta x$, which is proportional to the effective four-dimensional Schwarzschild radius of the wave packet, $R_{\rm S} \simeq G_4 \Delta p/c^3$, yielding
\begin{eqnarray} \label{GUP}
\Delta x \gtrsim \frac{\hbar}{2\Delta p} + \frac{2G_4}{c^3}\Delta p \, , 
\end{eqnarray} 
where $\alpha$ again denotes a numerical constant of order unity. 
Assuming, instead, that the GUP (\ref{GUP}) holds in a four-dimensional Universe, in place of the HUP (\ref{HUP}), we may expect a unification of the Compton and Schwarzschild lines, of the form
\begin{eqnarray} \label{CS}
R_{\rm C/S} \simeq \frac{\hbar}{2Mc} + \frac{2G_4}{c^2}Mc \, , 
\end{eqnarray} 
as predicted by the so-called BHUP correspondence, mentioned in the Introduction \cite{Carrr:2017aws,Lake:2018hyv,Carr:2022ndy,Lake:2016enn,Carr:2011pr,Carr:2015nqa,Carr:2014mya,Lake:2015pma,Singh:2017wrb,Singh:2017ipg,daSilva:2022xgx}. 
Combing these expressions with the arguments presented above yields even richer phenomenology: rather than simply restoring symmetry to the $(M,R)$ diagram higher dimensions, it may provide a way to unify the Compton and Schwarzschild lines, even in higher-dimensional spacetimes. 
Such an analysis lies outside the scope of the present, preliminary study, and is left to a future work.

\begin{figure}[h] \label{Fig.1}
\centering 
\includegraphics[width=16cm]{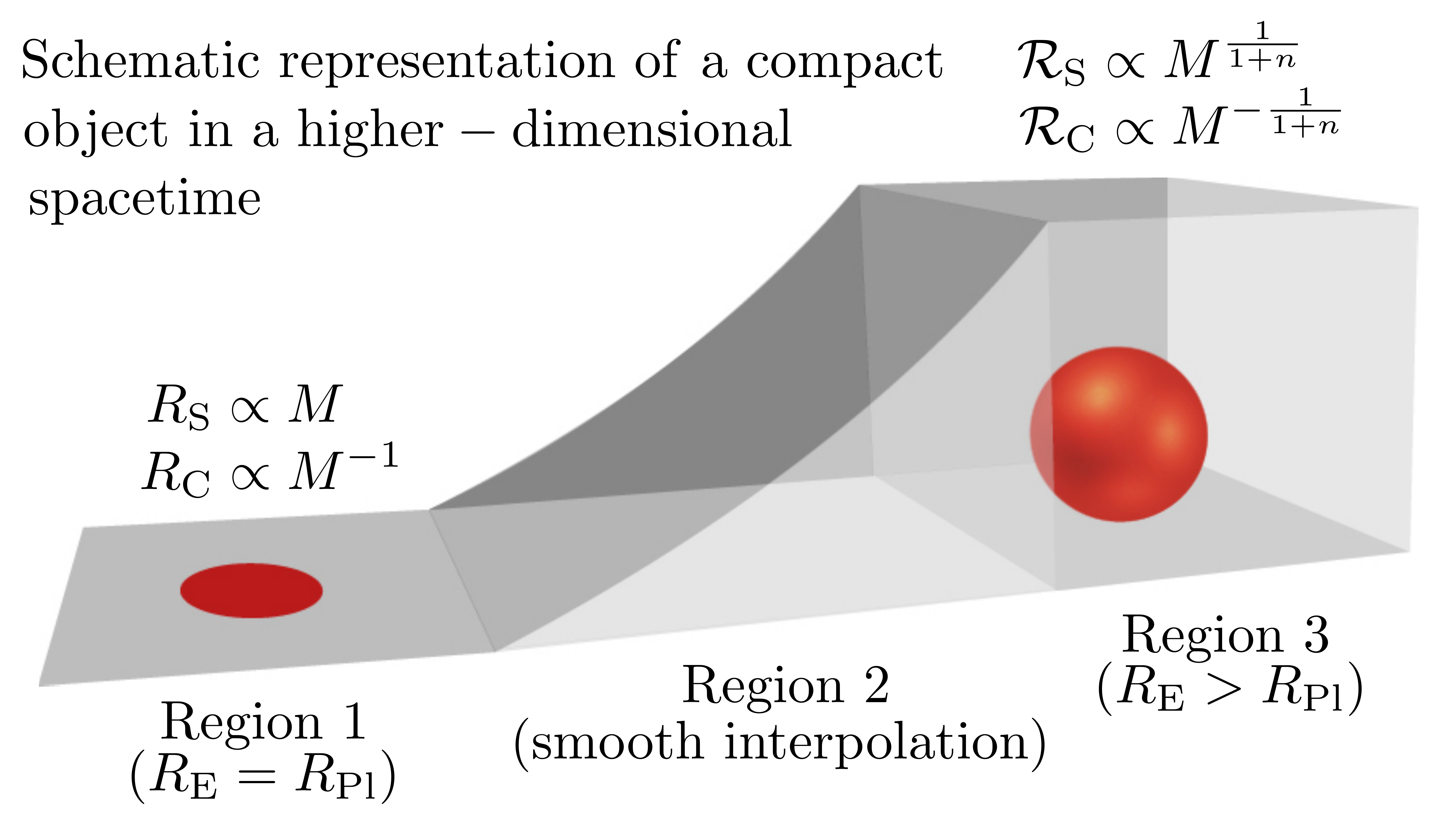}
\caption{Schematic illustration of the three-part universe in our gedanken experiment. To enable the schematic representation of $(3+n)$-dimensional space, neglecting the time dimension of the $(4+n)$-dimensional spacetime, the three large dimensions are depicted as a two-dimensional plane and the $n$ compact directions are depicted as a single extra dimension, extending into the $z$-direction of the diagram. Furthermore, since Planck-sized extra dimensions do not contribute correction terms, either to the higher-dimensional Schwarzschild radius, or to the Compton wavelength, we neglect them in this illustration. Hence, the region on the far left-hand side represents $(3+n)$-dimensional space, with $n$ dimensions compactified at the Planck scale, while the region on the far right-hand side represents a space with three large dimensions and $n$ extra dimensions, compactified on some scale $R_{\rm E} > R_{\rm Pl}$. The central region interpolates smoothly between the two, so that the gravitational radius of the compact body changes, according to the following scheme:
In region 1 (left), the extra dimensions are compactified at the (four-dimensional) Planck scale and both black holes and fundamental particles are effectively four-dimensional, even in the presence of the higher-dimensional space. 
In region 3 (right), the compactification radius is much larger than the Planck length and all objects are effectively $(4+n)$-dimensional, on scales smaller than the compactification radius. 
Conservation of energy implies that, whatever the mass-radius relation of the object in the first region, $R(M)$, its radius in the third region, $\mathcal{R}(M)$, must be larger: $\mathcal{R}(M) > R(M)$. 
This is due to the increased strength of the gravitational field in higher dimensions. 
For black holes, $R_{\rm S} \propto M$ in region 1 and $\mathcal{R}_{\rm S} \propto M^{\frac{1}{1+n}}$ in region 3. Applying the same logic to the gravitational radius of fundamental particles, $R_{\rm C} \propto M^{-1}$ in region 1, yielding $\mathcal{R}_{\rm C} \propto M^{-\frac{1}{1+n}}$ in region 3, due to the conservation of gravitational self-energy.}
\end{figure}

\begin{figure}[h] \label{Fig.2}
\centering 
	\includegraphics[width=12cm]{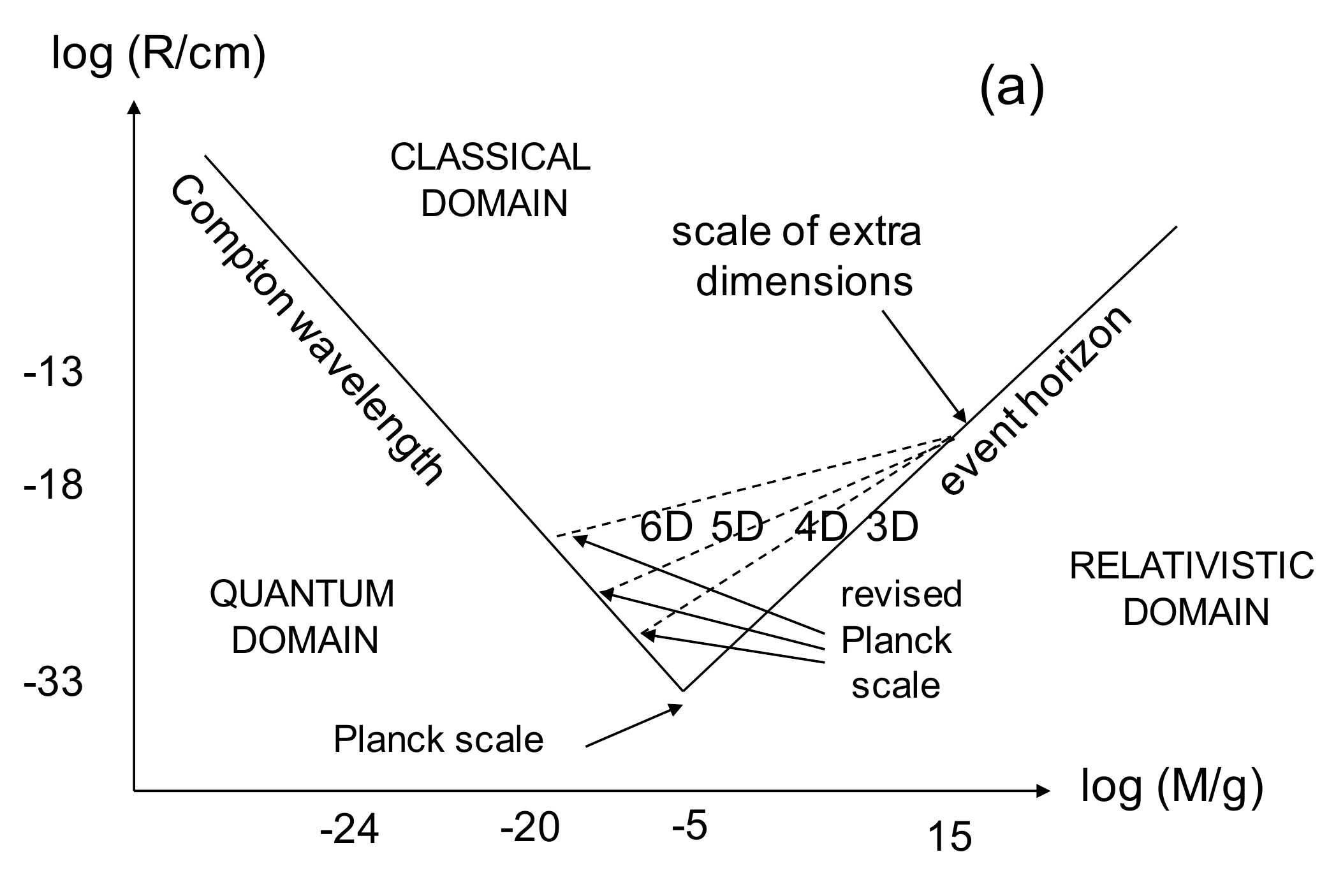}
	\includegraphics[width=12cm]{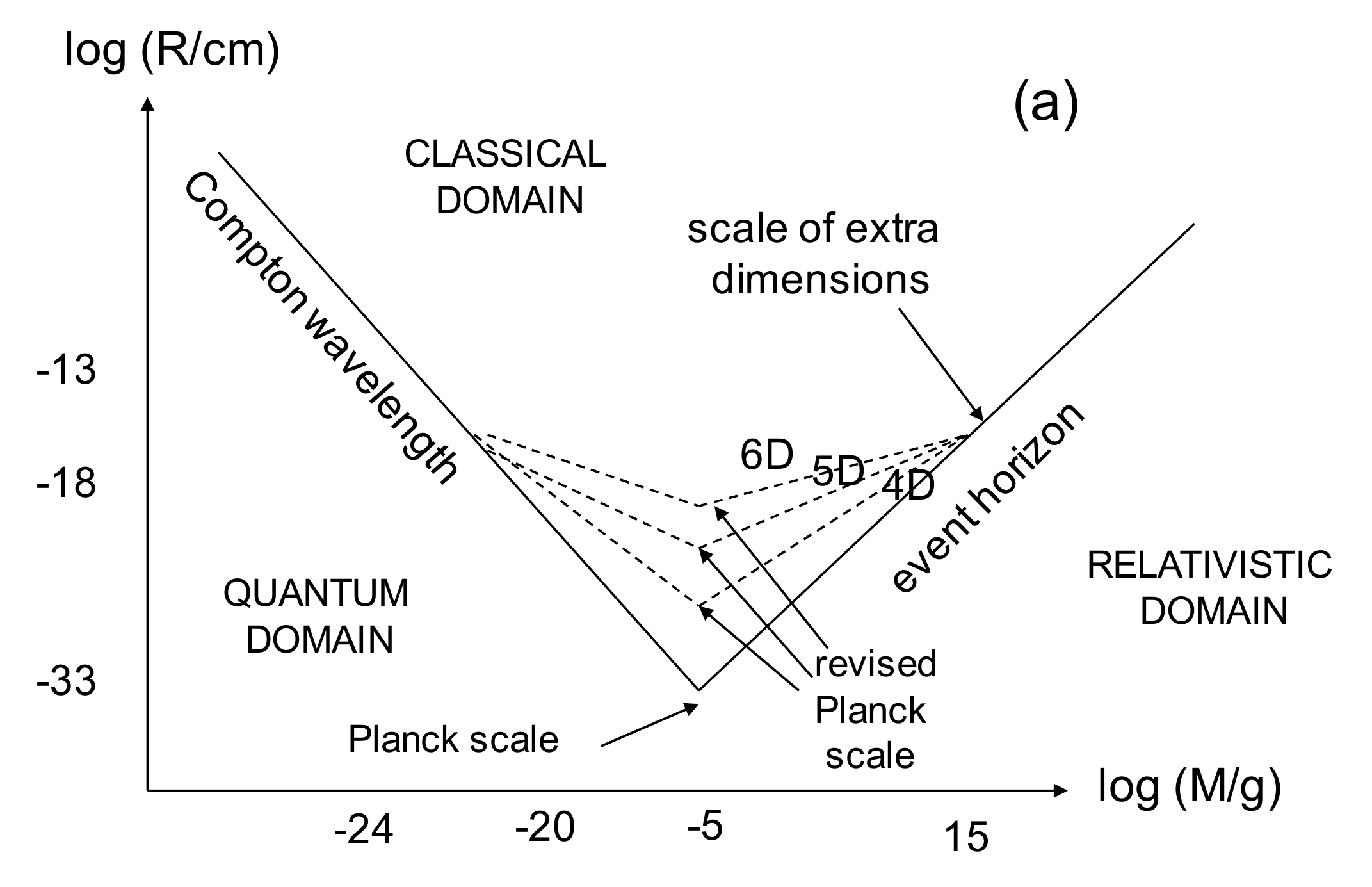}
\caption{Fig. 2a (top panel) shows the standard Compton line, $R_{\rm C} \propto M^{-1}$, and the Schwarzschild radius lines for $n=0$, $n=1$, $n=2$ and $n=3$. These lines intersect near the higher-dimensional Planck point, $(M,R) = ((M_{\rm Pl}^2M_{\rm E}^n)^{\frac{1}{2+n}},(R_{\rm Pl}^2R_{\rm E}^n)^{\frac{1}{2+n}}))$, where $R_{\rm Pl}$ and $M_{\rm Pl}$ denote the four-dimensional Planck scales, $R_{\rm E} > R_{\rm Pl}$ is the compactification radius, and $M_{\rm E} = \hbar/(R_{\rm E}c) < M_{\rm Pl}$ is the associated mass scale. The points of intersection are equivalent to the critical scales shown in Eq. (\ref{HD_Planck}), due to Eq. (\ref{G_4:G_4+n}). Fig. 2b (bottom panel) shows the modified scenario, in which we account for the increased self-gravity of the quantum particle in the presence of the extra dimensions, yielding $\mathcal{R}_{\rm C} \propto M^{-\frac{1}{1+n}}$. The Compton and Scwarzschild lines now intersect at the point $(M,R) = (M_{\rm Pl},R_{*})$, where $R_{*}$ is defined in Eq. (\ref{R_*}). The restored symmetry of the mass-radius diagram precludes the formation of black holes at TeV scales, even if large extra dimensions exist. These figures are reproduced from \cite{Lake:2018hyv}, with permission.}
\end{figure}
 
\section{Discussion} \label{Sec.3}

We have presented a simple gedanken experiment in a hypothetical spacetime with three macroscopic spatial dimensions and $n$ compact extra dimensions. 
The compactification radius was allowed to vary as a function of spatial position, in the four-dimensional submanifold, which is divided into three regions. 
In the first region, the extra dimensions are Planck-scale, while in the third they are compactified at a much larger radius. 
The second region, in which the compactification scale grows monotonically, interpolates smoothly between the other two. 
We considered a spherical compact object that traverses a path from region 1 to region 3, and imposed the conservation of gravitational self-energy. 

If the object is a black hole in the first region, with $R_{\rm S} \propto M$, energy conservation alone yields the correct expression for the higher-dimensional Schwarzschild radius, $\mathcal{R}_{\rm S} \propto M^{\frac{1}{1+n}}$, in the third. 
However, this procedure is agnostic to the mass-radius relation of the object.  
Hence, considering a fundamental particle instead of a black hole, we instead imposed the standard formula for the Compton wavelength, $R_{\rm C} \propto M^{-1}$, in the first region. 
Conservation of energy then implies the existence of a higher-dimensional Compton wavelength, $\mathcal{R}_{\rm C} \propto M^{-\frac{1}{1+n}}$, in the third region. 
Clearly, this relation must also hold for particles that have always been confined to region 3.

The new relation restores the symmetry between the Compton and Schwarzschild lines on the mass-radius diagram, in higher-dimensional spacetimes, and precludes the formation of black holes at TeV scales, even if large extra dimensions exist. 
We have shown how this follows, intuitively, as a direct consequence of the increased gravitational field strength at distances below the compactification scale. 
Combining these results with the usual, heuristic identification between the Compton wavelength and the minimum position uncertainty allowed by the Heisenberg uncertainty principle, $\Delta X \gtrsim \mathcal{R}_{\rm C}$ ($\Delta P \lesssim Mc$), suggests the existence of generalised, higher-dimensional uncertainty relations. 

Indeed, the possible dependence of the uncertainty relations on the dimensionality of the spacetime has already been explored in the literature, in the context of the so-called black hole-uncertainty principle (BHUP) correspondence \cite{Carrr:2017aws,Lake:2018hyv,Carr:2022ndy,Lake:2016enn}. 
If the usual uncertainty relation-Compton wavelength correspondence is still required to hold, in a higher-dimensional context, then the dimensional-dependence of the Compton wavelength is also (theoretically) necessary. 

The difference between this and previous work is that, here, we present a clear physical argument for {\it why} this change should occur, and show, explicitly, that the effects of self-gravitation on quantum wave packets are precisely those required to maintain the, up to now conjectured, higher-dimensional BHUP correspondence. 
This is also known as the Compton-Schwarzschild correspondence, in some of the previous literature \cite{Carr:2011pr,Carr:2015nqa,Carr:2014mya,Lake:2015pma,Singh:2017wrb,Singh:2017ipg,daSilva:2022xgx}.  

In the present, preliminary analysis, we assumed throughout that the gravitational potential of the compact sphere can be well approximated by the Newtonian regime. 
Though this is undoubtedly a limitation of the current work, we were still able to recover, to within numerical factors of order unity, the well-known expressions for relativistic objects, such as higher-dimensional black holes and neutron stars \cite{Burikham:2015nma,Burikham:2015sro}. 
This strongly suggests that the dimensionally-dimensional uncertainty relations, which we derive for self-gravitating wave packets, are robust, since the weak field approximation is undoubtedly valid for fundamental particles.

As extensions of the current analysis, we should consider relativistic corrections, as well as the incorporation of modified uncertainty principles, obtained from the quantum gravity literature, such as the generalised uncertainty principle (GUP) \cite{Adler:2001vs,Maziashvili:2005pp,Sakalli:2022xrb,Xiang:2009yq,Lake:2018zeg,Lake:2023lvh}, extended uncertainty principle (EUP), and extended generalised uncertainty principle (EGUP) \cite{Bolen:2004sq,Park:2007az,Bambi:2007ty}. 
Furthermore, in order to consistently incorporate the latter, we must also consider the conditions for the formation of gravitational bound states, in higher dimensions, in the presence of a positive cosmological constant \cite{Burikham:2015nma,Burikham:2015sro}.
 
Previous studies suggest that these modifications may give rise to a unified description of the Compton and Schwarzschild radii, linking the properties of black holes and fundamental particles in higher-dimensional scenarios \cite{Carrr:2017aws,Lake:2018hyv,Carr:2022ndy,Lake:2016enn}. 
The present work represents a small, preliminary step towards understanding the physical mechanism behind this potentially important correspondence, which may have important phenomenological implications for black holes, cosmology, and high-energy particle physics, beyond the non-production of PBH at TeV scales.

\section*{Acknowledgments}

Grateful thanks to Wansuree Massagram, for the artful rendering of Fig. 1, and Bernard Carr, for permission to reproduce Figs. 2, and for helpful discussions and suggestions during the preparation of the manuscript. 
ML acknowledges the Department of Physics and Materials Science, Faculty of Science, Chiang Mai University, for providing research facilities. 
AW would like to acknowledge partial support from the Center of Excellence in Quantum Technology, Faculty of Engineering, Chiang Mai University, and the NSRF via the Program Management Unit for Human Resources and Institutional Development, Research and Innovation (grant number B05F640218), National Higher Education Science Research and Innovation Policy Council. 
This work was supported by the Natural Science Foundation of Guangdong Province, grant no. 2021A1515010036. 



\end{document}